# LogDos: A Novel Logging-based DDoS Prevention Mechanism in Path Identifier-Based Information Centric Networks

Basheer Al-Duwairi[a,*], Öznur Özkasap[b], Ahmet Uysal[b], Ceren Kocaoğullar[b] and Kaan Yıldırım[b]

[a]*Department of Network Engineering & Security, Jordan University of Science & Technology, Irbid 22110, Jordan, Email: basheer@just.edu.jo*
[b]*Department of Computer Engineering, Koç University, Istanbul, Turkey, Email: oozkasap@ku.edu.tr*

---

**ABSTRACT**

Information Centric Networks (ICNs) have emerged in recent years as a new networking paradigm for the next-generation Internet. The primary goal of these networks is to provide effective mechanisms for content distribution and retrieval based on in-network content caching. The design of different ICN architectures addressed many of the security issues found in the traditional Internet. Therefore, allowing for a secure, reliable, and scalable communication over the Internet. However, recent research studies showed that these architectures are vulnerable to different types of DDoS attacks. In this paper, we propose a defense mechanism against distributed denial of service attacks (DDoS) in path-identifier based information centric networks. The proposed mechanism, called LogDos, performs GET Message logging based filtering and employs Bloom filter based logging to store incoming GET messages such that corresponding content messages are verified, while filtering packets originating from malicious hosts. We develop three versions of LogDos with varying levels of storage overhead at LogDos-enabled router. Extensive simulation experiments show that LogDos is very effective against DDoS attacks as it can filter more than 99.98 % of attack traffic in different attack scenarios while incurring acceptable storage overhead.

## 1. Introduction

Distributed denial of service (DDoS) attacks represent a major threat for traditional and future network architectures. Today's Internet has witnessed major DDoS attack incidents and attack rates have exceeded unprecedented limits. These attacks continue to grow in number and sophistication as evident by several attack incidents that were reported recently. For example, major DDoS attack targeted Github in March 2018 with aggregate traffic rate that exceeded 1.7 Tbps [27]. This came after the major incident of the well known DDoS attack on the DNS provider Dyn in October 2016 with a traffic rate around 1.2 Tbps. The Dyn attack originated from thousands of Internet of Things (IoT) devices controlled by the Mirai botnet [5]. DDoS attacks are not limited to traditional computer networks. They represent a major issue in different types of network architectures and systems such as software defined networks, cloud computing, edge/fog computing, vehicular networking, smart grid systems and named data networks. It is clear that preventing and mitigating these attacks is becoming a priority in network security in general, and there is a pressing need to address this issue in future Internet architectures as well.

Information Centric Networking (ICN) [29] has emerged as a future Internet communication model to address the need of having effective ways to provide content distribution and retrieval. Several network architectures were proposed in recent years to realize this communication model. This include Named Data Networks (NDN) [13] and Path-Identifier based ICN. NDN is based on Content Centric Networking (CCN) architecture, where content providers distribute their content to a large number of nodes (i.e., content stores), and content consumers obtain certain content by sending interest messages. A node that has the requested content replies back with a DATA message containing that content. Routing in these networks is different than that in the traditional IP networks. NDN relies on maintaining a Pending Interest Table (PIT) at each NDN router where outstanding interests and their arrival interface is stored such that content producers inject their traffic to be sent to the consumer traversing the same path followed by the interest messages. On the other hand, Path-ID based ICNs rely on using Path Identifiers (PIDs) to identify paths between network domains where these PIDs are added to interest messages along the path to the content provider such that content follow back the same path. Details of the two approaches are presented in Section 2.

Both NDN and PID-based ICNs are vulnerable to different types of DDoS attacks [24, 21, 22]. In NDN, PIT can be overloaded with flood of fake interest messages that would populate PITs with huge number of entries that correspond to fake interest messages resulting in a PIT overflow problem. On the other hand, in PID-based information centric networks the feasibility of DDoS attacks depends on attackers ability to learn path identifiers. In some PID-based architectures, PIDs are globally advertised and become public. Therefore, PIDs can be learned easily by attackers. However, in other PID-based architectures (e.g., [22]), it is recommended that PIDs are kept secret. Therefore, imposing additional overhead on attackers to learn PIDs. In both cases, an attacker controlling a set of compromised machines can know path identifiers using different techniques. Therefore, enabling the attacker to flood targeted system with huge amount of content messages along the learned paths.

In this paper, we propose a novel mechanism to address the problem of data flooding attack in PID-based ICNs, as an extension of our work presented in [4]. The proposed mechanism, called LogDos, performs `GET` Message logging based filtering that is similar to PID-based ICN architectures in the sense that it uses PIDs to perform inter-domain routing. However, we add `GET` message digests at ICN routers along the path to the content provider in a flavor similar to that used in PIT based NDN architecture aiming at filtering content packets that do not correspond to previously requested content. Storing `GET` message digits allows for verifying content messages while maintaining minimal storage at intermediate routers. The main contributions of the paper are as follows.

*Corresponding author.
ORCID(s):

- We propose LogDos as a novel mechanism to address the problem of data flooding attack in PID-based ICNs. To the best of our knowledge, LogDos is a unique hybrid approach that combines the best of NDN networks and PID-based ICNs. In order to reduce the storage overhead at LogDos-Enabled routers, we utilize Bloom Filters to store message digests in such a way that allows the verification of the corresponding data messages.

- We propose three variants of the LogDos scheme aiming at lowering the storage and processing overhead while achieving efficient filtering of attack traffic, namely the mechanisms based on 1) comprehensive-logging, 2) odd/even-logging, and 3) dynamic-logging. We demonstrate that the proposed LogDos mechanisms do not incur any communication overhead and are not affected by attackers ability to learn path IDs.

- We developed an event driven simulator [2] for modelling the path-ID based ICNs, consisting of modules for realistic Internet AS topology modeling, routing, packet representation and LogDos mechanisms. Extensive simulation experiments using large-scale network topologies and realistic attack scenarios are conducted for evaluating the LogDos mechanisms in terms of the effectiveness in preventing DDoS attacks in Path ID-based ICNs and a quantification of the storage overhead required to log packets.

- The experimental results show that LogDos mechanisms are very effective in countering data flooding attacks by adjusting bloom filter parameters at LogDos-enabled routers. Our findinda indicate that the overall attack rate observed at the victim did not exceed 0.3 Mbps, 7 Mbps, 50 Mbps, and 17 Mbps out of aggregate attack rate of 3 Gbps for comprehensive-logging, even-logging, odd-logging and dynamic-logging, respectively. This is by large far away from the rate observed at the victim in case of D-PID mechanism when attackers perform periodic PID learning.

The rest of this paper is organized as follows. In Section 2, we present the background information and attack model. Previous works are discussed in 3. Our mechanism is presented in Section 4. Evaluation is presented in Section 5. Finally, conclusion is presented in Section 6.

## 2. Background and Attack Model

Information Centric Networking (ICN) is emerging as a future Internet architecture to satisfy the current trend of using the Internet mainly for information dissemination as users are mostly interested in accessing information regardless of its location, and to address shortcomings of current host centric Internet and to provide better security and mobility support. ICN focuses on providing efficient ways to deliver content (information) to recipients. This includes the use of in-network storage and caching mechanisms, content naming approaches, and secure routing techniques. To this end, several information centric architectures have been proposed in recent years to meet these requirements. In this paper, we focus our discussion on two architectures that represent main approaches for content request and content routing in Information Centric Networks. These architectures are the Named data networks [13] and a path-ID based ICN architecture called CoLoR [20].

### 2.1. Named Data Networking

Two types of packets are used in NDN: `interest` messages and `DATA` messages. NDN routers forward interests towards the content producer responsible for the requested name, using name prefixes for routing. A content message includes a name, a payload and a digital signature computed by the content producer. Names are composed of one or more components, which have a hierarchical structure. In NDN notation,"/" separates name components (e.g., `just/cit/content1`). Content is delivered to consumers only upon explicit request. Each request corresponds to an `interest` message. Unlike content, interests are not signed. An `interest` message includes a name of requested content. In case of multiple content under a given name, optional control information can be carried within the interest to restrict desired content. Content signatures provide data origin authentication. Each NDN router maintains three data structures that include: (i) Pending Interest Table (PIT) to record incoming `interest` messages and their arrival interfaces (ii) The Forwarding Information Base (FIB) (iii) The Content Store (Cs). When an NDN router receives an `interest` message it checks the content store to see if the requested content is available in which case a content message is sent back through the same interface where the interest is received. Otherwise, the router adds an entry to its PIT along with the router interface where the packet is received. Then it relies on its FIB to forward the `interest` message to upstream router along the path towards the content provider. A `DATA` message containing the requested content is then forwarded back to the content consumer along the same path.

### 2.2. CoLoR Architecture

ColoR was proposed in [20] as an information centric networking architecture that is based on Path Identifier based routing. In CoLor, every content is assigned a location and application-independent and unique service identifier (SID). The SID is self-certified and remains valid as long as the content is available. Also, each node is assigned a unique self-certified and persistent node identifier (NID) that does not depend on the node location. Self certification in both SID and NID servers the purpose of authentication without the need of an external authority. The reachability information of content is maintained by a logical resource manger (RM) in each domain. Resource managers in different domains maintain certain relationship with each other (e.g., provider, customer, or peer) and each node within a domain knows the location of the resource manager of that domain. Routing within a domain is done based on intra routing algorithm. Routing between domains is done based on virtual path name which consists of list of path identifiers (PIDs) assigned to links between border routers.

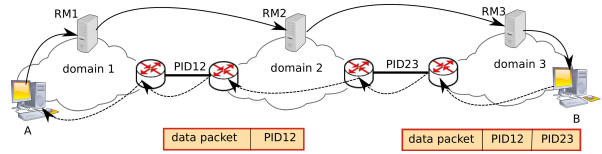

**Figure 1:** Operation of CoLoR information centric network architecture [20].

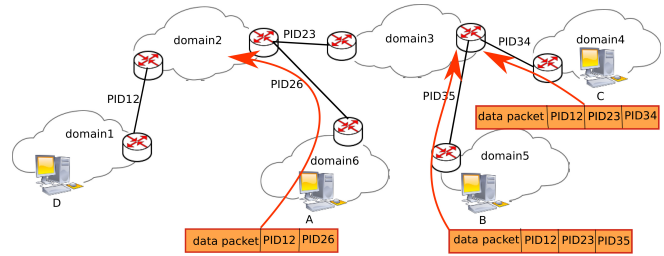

**Figure 2:** Data flooding attacks in CoLoR [22].

The operation of CoLoR is illustrated in Figure 1 which shows a network of three domains with a resource manger for each domain. A `GET` message is issued from node A to obtain certain piece of content. The message includes content SID and node's A NID (the consumer) get content with SID, and is forwarded to the local resource manager (RM1). Assuming that the request content is available at some provider B in domain 3, RM1 appends the path identifier of the path connecting domain 1 to domain 2 (i.e., PID12) which is along the path to the requested content. Once the message reaches domain 2, it is forwarded to RM2 in order to decide on the next hop towards the requested content. RM2 appends PID23 as the `GET` message and forwards it to the router connecting domain 2 with domain 3. Finally, RM3, forwards this message to node B which is located in domain 3. Since the list of PIDs along the path from consumer to provider is already encapsulate in the message, the provider (i.e., node B in this example) can send the content back along the same path. At each router along the backward path, path identifier connecting one domain to another is removed from the list.

### 2.3. DDoS attacks in Path Identifier-based Information Centric Networks

Data flooding attack is a type of DDoS attacks that can potentially target path identifier based information centric networks. In this attack, botnet machines are instructed to flood the target machine/network with high volume of data packets. Typically, in these networks, content is delivered from a provider to a consumer along the path followed by the `GET` message by including the list of received PIDs in the reply message as explained in Subsection 2.2. However, in an attack scenario, data packets are flooded towards the target directly (i.e., Not in a response to GET requests). This requires the attacker to know path identifiers (PIDs) used in the network.

Figure 2 shows a typical attack scenario. Nodes A, B, and C represent attack nodes that flood the target (node D) with data packets along the paths, PID34-PID23-PID12, PID35-PID23-PID12, PID26-PID12, respectively. It is clear that making path identifiers public allows attackers to easily launch data flooding attacks. Changing path identifiers dynamically as proposed in [21] can reduce the effect of data flooding attacks and imposes additional constraints on attackers. Also, it was shown in [22] that keeping PIDs secret can reduce the impact of these attacks. However, both strategies do not prevent DDoS attacks completely as attackers can use different techniques to learn path IDs in real-time. These include interest luring and botnet cooperation techniques [22].

### 2.4. Adversary Model

In the system model, we consider attackers who have the ability to flood the target system with high volume `DATA` messages from different attack machines clustered in multiple autonomous systems. Typically, such machines are part of a botnet controlled by the attacker. Before launching data flooding attacks, the list of path IDs along each attack path towards the victim should be learned. In [22], two mechanisms were proposed to learn path IDs assuming that they are kept secret. The first mechanism, called *interest luring*, relies on setting up a content server that hosts a popular content. Therefore, luring `GET` messages from consumers and learning PIDs found in these messages. The second mechanism, called *botnet cooperation*, relies on exchanging `GET` messages between different bot machines. Therefore, allowing the attacker to learn portion of PIDs along the paths followed by these messages. In addition, we assume that a malicious or compromised provider can still orchestrate the data flooding attack as it has full knowledge of the path represented by the list of PIDs contained in the received `GET` messages. In such a scenario, the attacker can collect PID information and craft fake data packets. Also, it is possible for the attacker to learn portion of PIDs through one or multiple compromised routers. A compromised router on the path would have a good knowledge of the path and hence can orchestrate the attack.

## 3. Previous Work on DDoS attacks in ICN

Distributed denial of service attacks are growing in number and level of sophistication. Defending against these attacks represent a top priority in today's Internet due to their devastating effect. Research efforts in this field have resulted in several techniques and approaches for attack detection, mitigation, prevention and traceback. For current Internet infrastructure, large number of DDoS countermeasures were proposed. For example, several techniques (e.g., Kill-bots [15], phalanx [9], JUST-Google [3], SkyShield [30] and Net-fence [19]) have been proposed to mitigate application-layer DDoS attacks. Other research (e.g., [16, 17]) focused on identifying the scan sources behind amplification DDoS attacks. In addition to many approaches (e.g., [14, 26, 12]) for filtering of DDoS attack traffic. With the increased interest in Information Centric Networking in recent years and the potential threat of DDoS attacks in these networks, several mechanisms have been proposed to address this important issue. The following subsections reviews main research in the area of DDoS attacks in ICN networks considering two main types: (i) Interest flooding attacks in NDN and (ii) Data flooding attacks in path identifier- based ICN.

### 3.1. Interest Flooding Attacks in NDN

Interest flooding attacks (IFAs) represent a major issue in named data networks (NDN) [34]. In these attacks, the Pending Interest Table (PIT) of a router is targeted by high request rate with spoofed content names. Several mechanisms were proposed to address this type of attacks. The mechanism proposed in [1] relies on threshold values for request rate, cache hit ratio, contents rating, and publishers ratings to detect cache pollution attacks. It also devises some defense mechanisms that depend on the attack scenario, where cache content is replaced based on the rating of the content and its usage pattern. In [23], a detection mechanism based on the statistical hypothesis testing theory was proposed where the detector employs threshold values that do not depend on users' behavior on each interface of a router within the protected network. The paper did not address advanced attack types such as those described in [28] and did not provide mitigation of the attack once detected. In [31], the importance of implementing Interest NACK into content centric networks was analyzed and evaluated as a potential solution to interest flooding attacks in these networks because it helps in evacuating interest tables of ICN routers without necessarily waiting for each interest until it expires. Poseidon was proposed in [8] as a framework for interest flooding attack detection and mitigation by relying on both local metrics and collaborative techniques for early detection of interest flooding attacks. InterestFence was proposed in [10] as a lightweight IFA detection and filtering mechanism based by the content server rather than routers. In [18], bloom filters are used to reduce memory cost at NDN routers.

### 3.2. Data Flooding Attacks in Path Identifier-Based ICN

DDoS attacks in information centric networks that use path-identifiers (PIDs) for inter-domain routing were investigated in [21] and [22]. Building on "*off by default*" approach [6] and "*capability-based*" approaches [32, 25], the work presented in [21] represents the first mechanism, called *D-PID*, to counter data flooding attacks in path-based ICNs. The paper shows that dynamically changing path IDs between domains can potentially prevent attackers who have the ability to learn PIDs from flooding a targeted system by irrelevant content packets. Moreover, the work presented in [22] emphasized that keeping PIDs secret is much better than making them public from security point of view and reduces attackers ability to perform DDoS attacks unless they use some advanced techniques by which they learn PIDs. The work presented in [21] and [22] represent a milestone for evaluating PIDs based ICNs resilience for DDoS attacks by using dynamic PIDs and by keeping PIDs secret. However, the two approaches suffer from major problems. On one hand, changing PIDs dynamically would result in an unnecessary overhead and does not prevent DDoS attacks in case that attacker learn PIDs dynamically. On the other hand, there are several mechanisms that attackers can use to learn PIDs even if they are kept secret.

The proposed LogDos mechanism advances the state-of-the-art in defending against DDoS attacks in path-ID based ICNs. Mainly, LogDos addresses the problems of D-PID mechanism proposed in [21] by (i) avoiding the communication overhead that would result from periodic exchange of path IDs between adjacent routers (ii) eliminating the need to keep path IDs secret regardless of the attackers ability to learn path IDs using interest luring and botnet cooperation mechanisms described in [22] (iii) Dealing with attack scenarios described in Subsection 2.4 which include learning path IDs through compromised routers or compromised content providers.

## 4. Proposed LogDos Mechanism

We propose an effective mechanism, named LogDos, that utilizes GET message logging based filtering to prevent data flooding attacks in path identifier based information centric networks. The main idea of the LogDos mechanism is to log some information contained in the GET messages passing through ICN routers such that these routers can verify corresponding response DATA messages. Here, LogDos leverages the routing symmetry feature in path-ID based ICNs, which allows an LogDos-enabled router that performed logging of a given GET message to verify the corresponding DATA message. For example, with reference to Figure 3, when a node X which is a content consumer sends a GET message to obtain certain content and that content is available at node Y (i.e., the content provider). Then, based on the routing mechanism of path-ID based ICNs, each LogDos-enabled router along the path from X to Y will append to the GET message the path ID assigned to the link between itself and the next router.

In this example, Router A will append ID1, router B will append ID2, router C will append ID3 and Router D will append ID4. At the same time each router, along the path will log the GET message. Therefore, when the GET message reaches Y it will contain full path information with list of path IDs: ID1-ID2-ID3-ID4. Node Y responds with a DATA message containing the requested content and send it along the same path but in reverse order. Each router along the path back to node X would be able to verify the validity of the DATA packet. Attack packets originating from Attacker1 and Attacker 2 and targeting the consumer network along paths ID6-ID1 and ID5-ID3-ID2-ID1 can be filtered directly because they do not correspond to previously logged GET messages. In the following subsections, we present three versions of LogDos with varying levels of storage overhead at routers.

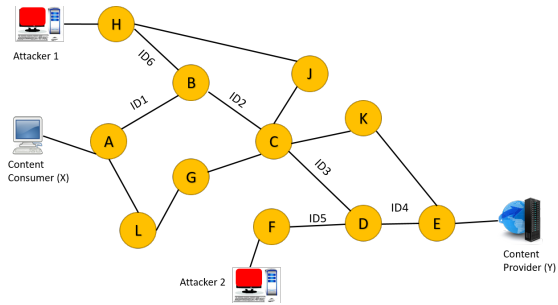

**Figure 3:** Illustrative example of the proposed mechanism

### 4.1. Comprehensive Logging:

In order to reduce the storage overhead associated with GET messages' logging, we propose to use Bloom filters [7] to log GET message digests instead of logging the message itself or part of it. Description of Bloom Filters is provided in Appendix A. In the proposed mechanism, we use the SID and the list of PIDs in the GET message as input for the hash functions of the bloom filter as they uniquely identify each GET message. In this paper, we refer to logging of all GET messages at each router as *Comprehensive Logging*. In theory, if there exist *n* LogDos-enabled routers along the path from an attacker to a victim, then the probability that an attack packet reaches the victim is given by Equation 1.

$$Pr(attack) = p^n \qquad (1)$$

Where *p* is the bloom filter's false positive probability given by equation 4 in Appendix A. It is to be mentioned that logging is done at LogDos-enabled router where the GET message leaves a domain, while verification is done at ICN routers where the response message enters a domain. This way, we ensure that logging and verification are done at the same router, and that a GET message is logged only one time in each domain. Algorithm 1 shows the comprehensive logging based filtering at an LogDos-enabled router, where the router logs all GET messages coming through an external interface.

It is to be mentioned that since Bloom filters do not allow delete operation, the false positive rate increases with the number of items that are inserted in Bloom filter to the extent that it is considered saturate. In order to deal with this issue, we propose that each LogDos-enabled router would

---

**Algorithm 1:** Comprehensive Logging based filtering

---
**Input** : Incoming Packet (P)
**Output**: Filtering decision

**1 for** *each incoming packet (P)* **do**
**2**     **if** *(P is a* GET *message) and (P is coming through an external interface)* **then**
       • Insert P in R's bloom filter
       • Allow P
**3**     **end**
**4**     **if** *(P is a* DATA *Message) and (P will be forwarded through an external interface)* **then**
**5**        **if** *(P is in R's Bloom Filter)* **then**
          • Allow P
**6**        **else**
          • reject P
**7**        **end**
**8**     **end**
**9 end**

---

maintain at least two Bloom filters such that when the number of inserted items in the first Bloom filter reaches certain value that corresponds to the targeted false positive rate, the router switches to the second bloom filter. A response DATA message would be validated by checking the two bloom filters. Once the second bloom filter reaches its limit, the first bloom filter is reset and new GET messages are inserted in it.

### 4.2. Odd/Even Logging:

Logging overhead can be reduced significantly by performing logging of GET messages at certain LogDoS-enabled routers along the path from content consumer to content provider, such that verification of corresponding DATA messages can be done at the same routers. To achieve this goal, we propose *Odd/Even Logging*. In Odd Logging, an LogDos-enabled router logs GET messages that have odd number of PIDs. The corresponding DATA messages are supposed to have the same SID and PIDs. Therefore, a LogDos-enabled router is supposed to verify DATA messages that have odd number of PIDs. Even logging is similar to Odd logging except that routers perform logging of GET messages that contain even number of PIDs and verification is done for DATA messages that contain even number of PIDs. For example, assuming odd logging, the GET messages forwarded from node X to node Y along the path A, B, C, D, E will be logged only at B and D. Verification of corresponding DATA messages will be done by the same routers since they are supposed to carry odd number of PIDs by the time it reaches them. Algorithm 2 shows the odd logging based filtering at a LogDos-enabled router. A Router only logs GET messages containing an odd number of PIDs and coming through an external interface (i.e., at the router where the GET message enters the domain).

---

**Algorithm 2:** Odd Logging based Filtering

---

**Input** : Incoming Packet (P)
**Output:** Filtering decision

---

**1 for** *each incoming packet (P)* **do**
**2**     **if** *(P is a* GET *message) and (P is coming through an external interface) and (P.list of PIDs is odd)* **then**
        • Insert P in R's bloom filter

        • Allow P

**3**     **end**
**4**     **if** *(P is a* DATA *Message) and (P will be forwarded through an external interface) and (P. list of PIDS is odd)* **then**
**5**        **if** *(P is in R's Bloom Filter)* **then**
           • Allow P

**6**        **else**
           • reject P

**7**        **end**
**8**     **end**
**9 end**

---

In this case, then the probability that an attack packet reaches the victim, assuming that attack path consists of $n$ LogDos-enabled routers, is given by Equation 2.

$$Pr(attack) = p^{\lfloor n/2 \rfloor} \qquad (2)$$

Even Logging is exactly similar to Odd Logging except that LogDos-enabled routers log GET message that contain even number of PIDs. In this case, then the probability that an attack packet reaches the victim, assuming that attack path consists of $n$ LogDos-enabled routers, is given by Equation 3.

$$Pr(attack) = p^{\lceil n/2 \rceil} \qquad (3)$$

### 4.3. Dynamic Logging:

In this version of the proposed mechanism, routers perform packet logging and validation during certain intervals that are decided by each router dynamically. Rather than logging GET messages continuously as in the previous two cases, logging can be done *on* and *off* as packet logging would not be required when there is no attack. The proposed dynamic logging is done as follows. Each router performs logging for certain period of time $T$, then goes silent (i.e., without logging) for another period of time $S$. Initially each router sets its logging interval to $T_0$ and extends it by the same value each time an attack is detected by observing that the number of invalid DATA messages exceeds certain threshold during that interval. Validation of DATA packets is done during the same time interval shifted by small amount of time $\delta$ in order to account for round trip delay between forwarding a GET message and receiving the corresponding content message. Algorithm 3 shows the details of dynamic logging based filtering performed by each router.

---

**Algorithm 3:** Dynamic Logging based Filtering

---

**Input** : Initial time $t_0$, Silent period $T$, Initial logging duration $=T_0$
**Output:** Filtering decision for each incoming packet

---

**1** $c = 0$
**2 for** *each incoming Packet(P)* **do**
**3**     t= P.arrival_time;
**4**     **if** *(($t_0 < t < t_0 + T$) and (P is a* GET *message) and (P is coming through an external interface))* **then**
        • Insert P in R's bloom filter

        • Allow P

**5**     **end**
**6**     **if** *($t_0 + \delta < t < t_0 + T + \delta$)) and (P is a Response Message) and (P will be forwarded through an external interface)* **then**
**7**        **if** *(P is in R's Bloom Filter)* **then**
           • Allow P

**8**        **else**
           • reject P

           • $c ++$

**9**        **end**
**10**     **end**
**11**     **if** *($t > t_0 + T + \delta$)* **then**
**12**        **if** *(c > Threshold)* **then**
           • $T = T + T_0$

           • $c = 0$

**13**        **else**
           • $t_0 = t_0 + T + S$

           • $T = T_0$

**14**        **end**
**15**     **end**
**16 end**

---

As an illustrative example, let us consider the scenario shown in Figure 4. In this example, packet logging starts at time $t_0$ and logging interval $T$ is set to the initial value of $T_0$. DATA packet validation (i.e., checking that corresponding GET messages have their digests stored in router's Bloom filter) starts at time $t_0 + \delta$ in order to account for round trip delay. Assuming that the number of invalid response packets during the logging interval exceeds certain threshold value. Then an attack is detected, and the logging interval is extended by another $T_0$ (i.e., $T = T + T_0$). Assuming that no attack is detected during the extended logging period, then the router switches to the silent mode (i.e., no logging) for a period of $S$ seconds. After that the start time $t_0$ is update to $t_0 + T + S$ and the same process repeats.

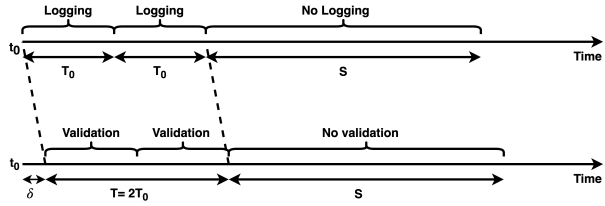

**Figure 4:** Dynamic logging based filtering example

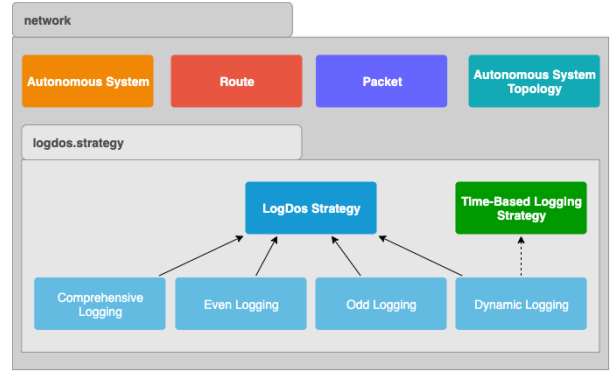

**Figure 5:** Simulator block diagram

## 5. Evaluation

In order to evaluate the effectiveness of our proposed LogDos scheme in preventing data flooding attacks in path-ID based ICNs, we developed an event driven simulator and conducted comprehensive simulation experiments. Subsection 5.1 discusses the simulator that we developed and the experimental setup. Subsection 5.2 discusses the simulation results.

### 5.1. Simulator and Setup

As there exists no simulator for modelling the path-ID based ICNs, we developed an event driven simulator [2] using an object-oriented approach and utilizing parallelism to optimize time performance. As depicted in Figure 5, our simulator consists of modules for realistic Internet AS topology modeling, routing, packet representation and LogDos strategies.

The simulation imports a predefined topology to create the autonomous systems and the edges connecting the routers. Once the topology is constructed, the simulator selects a victim and attackers randomly. Each simulation configuration is run multiple times to compensate for possible edge cases that have negligible probability but are possible. After designating a victim and all the attackers, the simulator finds the shortest paths from the attackers to the victim to send attack packets through. In order to simulate different logging strategies of LogDos, all autonomous systems have a logging strategy assigned to them.

Even though logging strategies differ, logging a packet means putting it to bloom filters existing on each AS. To achieve this, we employed Guava's Bloom Filter implementation [11]. Before initiating the attacks, each bloom filter is filled with random packets to imitate regular packet traffic. Otherwise, the logging strategies would be deceptively successful in catching attack packets. Simulating dynamic logging strategy requires a notion time. To represent time, our simulator uses a customized tick based timing method where all attack packets are assigned a random tick in which they will be sent. The delta between the two ticks are used to set router states appropriately before logging occurs. Our simulations utilize AS-level topology of current Internet adopted from [22] with its characteristics shown in Table 1.

**Table 1**
AS topology characteristics

| | |
|---|---|
| Number of ASes | 49,608 |
| Number of Edges | 212,543 |
| Transient ASes | 8,001 |
| Core ASes | 41,607 |
| Average attack path length | 3.86 |
| Standard deviation of path length | 0.04 |

### 5.2. Simulation Results

In this section, we evaluate the effectiveness of the proposed LogDos mechanism in defending against data flooding attacks by considering the three versions of LogDos. Then we discuss the storage overhead incurred by LogDos. Finally, we compare our LogDos to D-PID mechanism.

#### 5.2.1. Countering Data Flooding Attacks

In the first set of experiments, our main objective is to evaluate the effectiveness of the proposed LogDos mechanism in defending against data flooding attacks. The goal is to evaluate attack traffic rate observed at the victim under different simulation scenarios for each version of LogDos. Starting by Comprehensive Logging, Figure 6 depicts the overall attack rate observed at the victim for aggregate attack rates of 1 Gbps, 2 Gbps, and 3 Gbps, respectively. In each case, we vary the number of attacking ASes and distribute the overall attack traffic among them. Attack ASes represent autonomous systems that contain bot machines controlled by the attacker. In our simulation experiments, we set the number of attacking ASes to 100, 200, 500, 1000, and 2000, respectively. These ASes are selected randomly in each simulation run. In practice, bots are usually clustered in limited number of ASes [5]. Therefore, our choices of number of attack ASes reflects realistic attack scenarios. As shown in Figure 6, attack traffic rate at the victim increases by increasing aggregate attack traffic rate from 1 Gbps to 3 Gbps. However, it did not exceed 0.3 Mbps which is a very small value and does not have great impact on the victim. By adjusting Bloom filter parameters to achieve low false positive rates, the attack rate at the victim can be minimized significantly to a value that does not exceed $1 \cdot 10^{-2}$ Mbps. This is due to the fact that attack traffic has very low chance to pass through multiple routers along the path towards the victim.

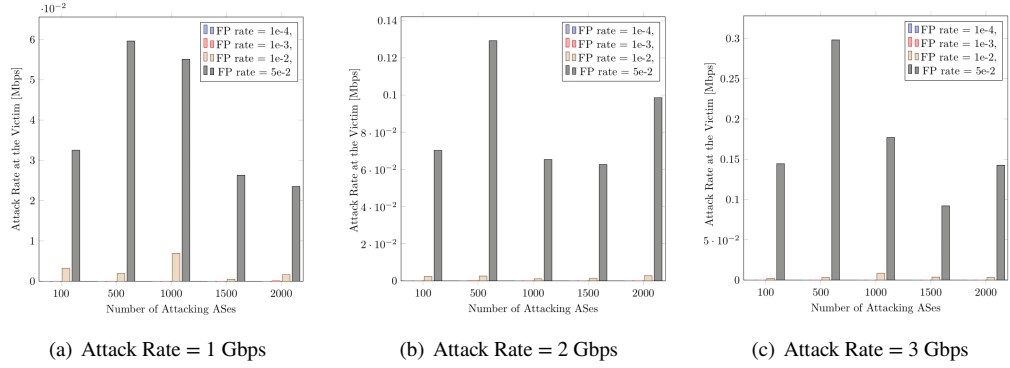

(a) Attack Rate = 1 Gbps       (b) Attack Rate = 2 Gbps       (c) Attack Rate = 3 Gbps

**Figure 6:** Attack rate at the victim- Comprehensive logging

In Section 4, we argued that it is possible to reduce the storage overhead at LogDos-enabled routers without sacrificing the ability to filter attack traffic by employing Odd, Even, or Dynamic Logging techniques. A `GET` message will be logged by $\lfloor n/2 \rfloor$ routers in Odd Logging and by $\lceil n/2 \rceil$ routers in Even Logging, where n is the number of LogDos-Enabled routers along the path. Hence, the probability of filtering an attack packet is greater in case of Even Logging compared to Odd Logging. This explains why Even logging is more effective than Odd Logging in filtering attack traffic as shown in Figures 7 and 8, respectively. Similar to comprehensive logging, in Even Logging and Odd logging, the attack traffic rate at the victim increases by increasing aggregate attack traffic rate. However, it did not exceed 7 Mbps for an attack rate of 3 Gbps and relatively a large false positive rate of 0.05 in case of Even Logging. On the other hand, it also did not exceed 50 Mbps for an attack rate of 3 Gbps and relatively a large false positive rate of 0.05 in case of Odd Logging. Simulation results show that, the overall attack traffic rate observed at the victim drops sharply for low false positive rates. This comes at the cost of increasing bloom filter size at LogDos-enabled routers as we will discuss in subsection 5.2.2.

In Dynamic Logging, LogDos-enabled routers perform `GET` message logging and `DATA` message validation in an ad hoc manner. Meaning that each router decides by itself when to enable/disable LogDos mechanism. Therefore, along a given attack path composed of $n$ routers, there would be time intervals when some routers are in the silent mode (i.e., LogDos is disabled), which means that not all routers are expected to enable LogDos simultaneously. This explains why Dynamic Logging is not as effective as Comprehensive Log-ging in terms of filtering attack traffic. However, it is comparable in performance to Even and Odd Logging as shown in Figure 9. Similar to Comprehensive and Even/Odd Logging, the overall attack rate observed at the victim increases by increasing aggregate attack traffic rate. However, it remains within acceptable range that does not have great impact on the victim. Overall, the proposed LogDos mechanism is very effective against data flooding attacks as it can filter more than 99.98 % of attack traffic in different attack scenarios.

### 5.2.2. Storage Overhead

The size of the Bloom filter required at each LogDos-enabled router for `GET` message logging depends on the overall `GET` messages arrival rate at the router. This in turn depends on the number of router interfaces and their link rates. In named data networking and according to [33], the overall number of PIT entries is estimated to be around 2M for a medium scale router with 36 interfaces each with 10 Gbps. We can project similar numbers for the number of entries to be logged by LogDos-enabled router. However, bloom filter size should be selected in such a way that the desired false positive rate is achieved. There is a trade-off between bloom filter size and false positive rate. Figure 10 shows the relationship between bloom filter size and false positive rate for different numbers of `GET` messages to be logged at a given router ranging from 0.5M to 2M. In practice, these numbers depend on different factors such as number of router interfaces, link rates, and router location.

For example, for a medium scale router that is expected to log 2M `GET` messages, the required bloom filter size is about 120 MB with false positive rate of $1 \cdot 10^{-4}$ assuming that the number of bloom filter hash functions equal to 3. For higher false positive rates and larger $k$ values, the required bloom filter size becomes smaller. In general, It is clear that the storage overhead incurred by LogDos mechanism is not significant especially when applying Odd/Even logging or Dynamic logging because it is not required to log every `GET` message.

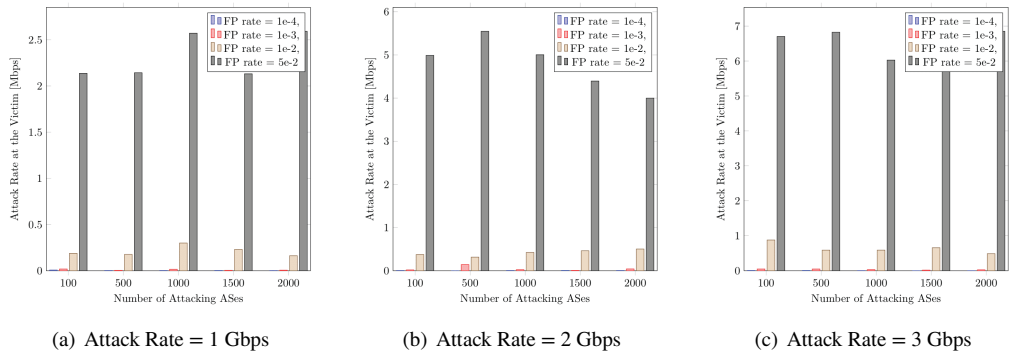

**Figure 7:** Attack rate at the victim- Even logging

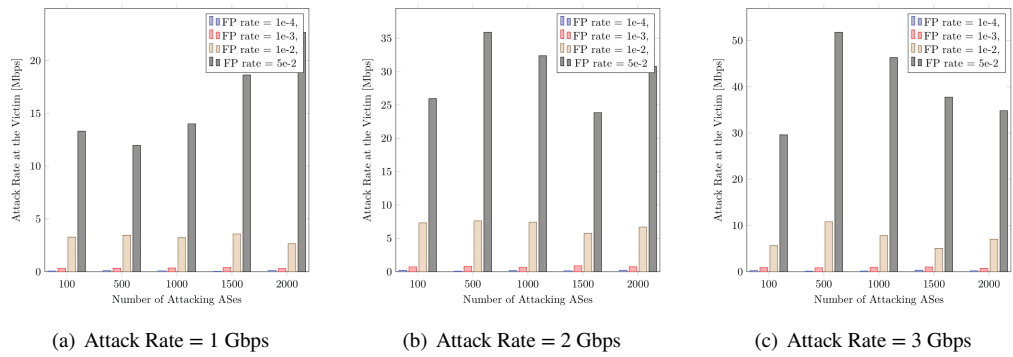

**Figure 8:** Attack rate at the victim- Odd logging

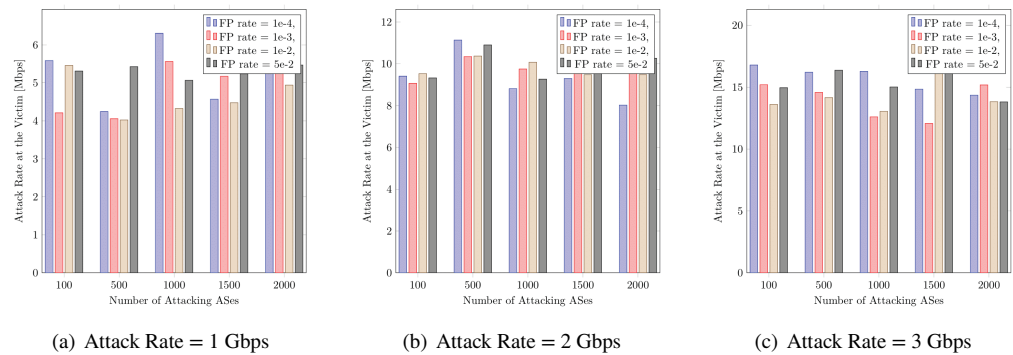

**Figure 9:** Attack rate at the victim- Dynamic logging

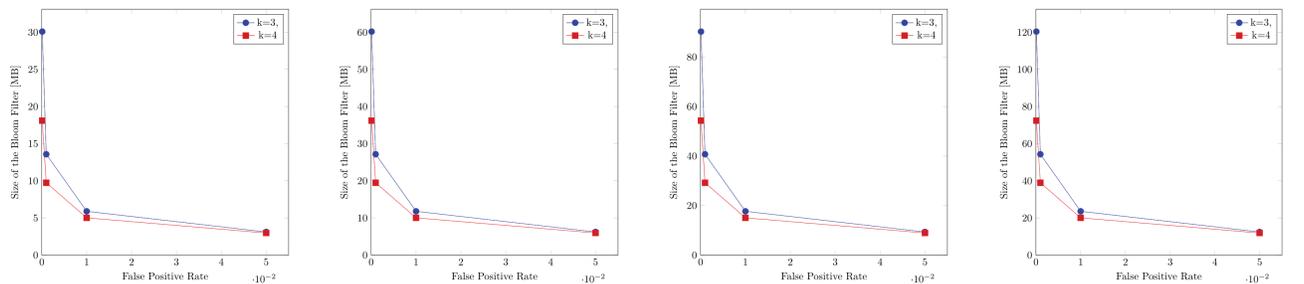

**Figure 10:** Storage Overhead at LogDos-Enabled Routers

### 5.2.3. Comparison with D-PID mechanism

A major problem in path-based ICNs is that an attacker who learns path IDs can flood the victim with DATA packets along the learned paths. In [21], it is shown that this can be achieved using two major techniques. The first technique is called *GET luring* where a node under attacker's control registers normal content names aiming at luring GET messages from content consumers. Therefore, learning path IDs included in these messages. The second technique is called *botnet cooperation* where bots controlled by the attacker are configured to exchange GET messages among each others. In order to prevent attacker from performing data flooding attacks even if they were able to learn path IDs (or part of them), a mechanism called D-PID was proposed in [21]. D-PID prevents DDoS attacks by changing path IDs in a dynamic way. In D-PID, every router negotiates a new set of path IDs (PIDs) with its neighbours every T seconds, where T is the PID update period. The major weaknesses of D-PID are that it incurs communication overhead and it can not prevent attacks that are based on periodically and continuously learning PIDs. In contrast to that our LogDos mechanism does not rely on keeping PIDs secret or on updating them dynamically.

In order to evaluate the overall attack rate at the victim in D-PID mechanism in case the attacker has the ability to learn PIDs periodically, we performed set of experiments assuming that learning full paths from attackers to the victim (i.e., list of PIDs along the path from attacker to victim) is a Poisson process with rate $\lambda$ and that Path ID update period is set to $T$ seconds. Also, we assume that the number of ASes controlled by the attacker = 100 and that attack rate per attacking AS = 20 Mbps. Figure 11 shows the aggregate attack rate at the victim for PID update periods 60, 120, and 240 seconds, and for different values of $\lambda$ (the rate of learning full attack paths). It can be seen that, D-PID fails to defend against DDoS attacks in case the attacker is able to quickly and periodically learn PIDs and identify full attack paths. Even with relatively small PID update period of 60 s, an attacker can flood the victim with high volume attack traffic reaching 2 Gbps for $\lambda = 8$. Figure 12 compares D-PID to Even Logging for false positive rates of 0.05, 0.1 and 0.2 respectively, while fixing the PID update period to 240 seconds. It is clear that that attack rate observed at the victim is very small even for a relatively large false positive rate. With reference to the results discussed in Subsection 5.2.1, it can be seen that our LogDos mechanism in its three versions outperforms D-PID mechanism in case attackers are able to learn PIDs dynamically and periodically.

## 6. Conclusion

Addressing the significance of dealing with DDoS attacks in future Internet architectures, we propose LogDos mechanism to counter these attacks in path identifier-based information centric networks. LogDos utilizes Bloom filter based logging to store incoming GET messages and verify them by storing a message only once per domain. We propose three versions of LogDos, namely Comprehensive, Odd/Even and Dynamic logging, with the objective of reducing the storage and processing overhead while achieving efficient filtering of attack traffic. We developed simulation models of LogDos methods, and conducted extensive simulations using large-scale network topologies and realistic attack scenarios. Simulation results show that different versions of LogDos are very effective in countering data flooding attacks by adjusting bloom filter parameters at LogDos-enabled routers. For example, overall attack rate observed at the victim did not exceed 0.3 Mbps, 7 Mbps, 50 Mbps, and 17 Mbps out of aggregate attack rate of 3 Gbps for Comprehensive Logging, Even logging, Odd logging and Dynamic Logging, respectively. This is by large far away from the rate observed at the victim in case of D-PID mechanism when attackers perform periodic PID learning. At the same time, the storage overhead incurred by LogDos is relatively small.

## Acknowledgement

The work presented in this paper is the outcome of the research visit of Dr. Al-Duwairi to Koç University supported by the Deanship of Research at Jordan University of Science and Technology under grant No. 20190187.

## A. Bloom Filters

A Bloom filter is a data structure for representing a set of $n$ elements (also called keys) to support membership queries. The idea, as illustrated in Fig. 13, is to allocate a vector $R$ of $m$ bits, initially all set to 0, and then choose $k$ independent hash functions, each with range $\{1,...,m\}$. For each element, $A$, the bits at positions $H_1(A)$, $H_2(A)$, ..., $H_k(A)$ in $R$ are set to 1. (A particular bit might be set to 1 multiple times.) Given a query for $B$, we check the bits at positions $H_1(B)$, $H_2(B)$, ..., $H_k(B)$. If any of them is 0, then certainly $B$ is not inserted in the filter. Otherwise, we conjecture that $B$ is inserted in the filter although there is a certain probability that we are wrong. This is called a "false positive". The parameters $k$ and $m$ should be chosen such that the probability of a false positive is acceptable. The false positive rate of a Bloom filter is given by the following equation:

$$p = (1 - (1 - \frac{1}{m})^{kj})^k \approx (1 - e^{\frac{kj}{m}})^k \qquad (4)$$

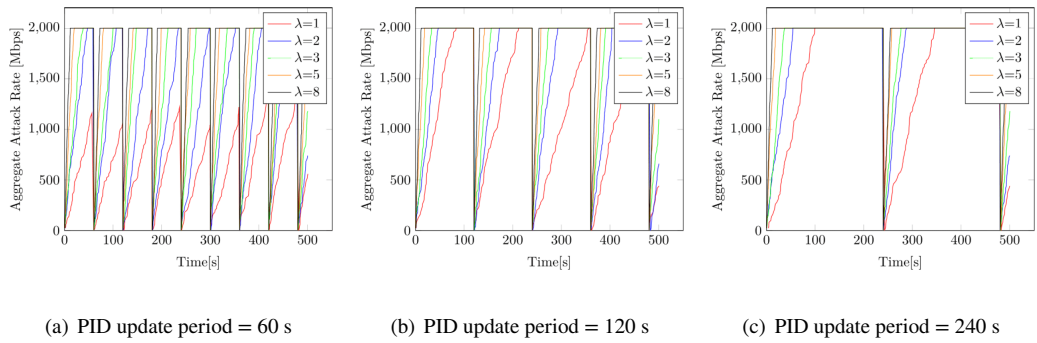

(a) PID update period = 60 s    (b) PID update period = 120 s    (c) PID update period = 240 s

**Figure 11:** Attack rate at the victim- D-PID

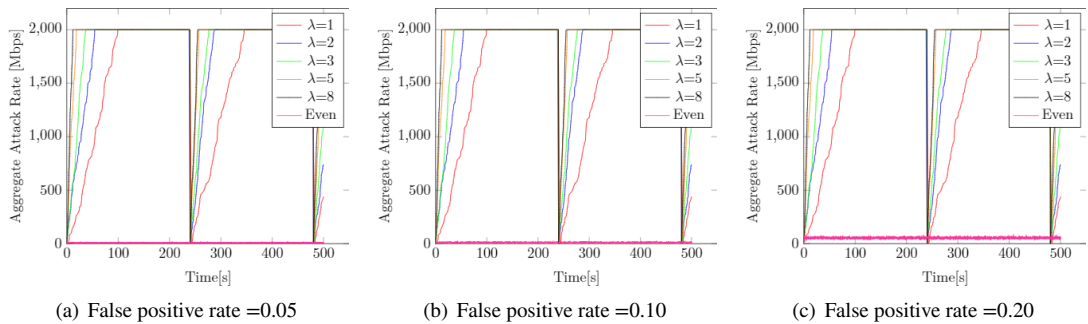

(a) False positive rate =0.05    (b) False positive rate =0.10    (c) False positive rate =0.20

**Figure 12:** Attack rate at the victim- D-PID vs. Even Logging

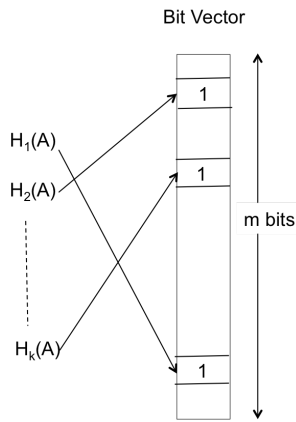

**Figure 13:** A Bloom filter with $k$ hash functions

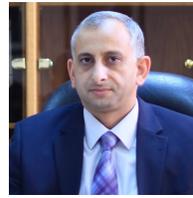

**Basheer Al-Duwairi** is an Associate Professor at the Department of Network Engineering and Security at Jordan University of Science & Technology. He received his B.S. in electrical and computer engineering from Jordan University of Science and University (JUST) in 1999, and his M.S. and PhD in computer engineering from Iowa State University, Ames, IA in 2002 and 2005, respectively. Over the past 15 years, Dr. Al-Duwairi research interests are in the area of network security focusing mainly on developing efficient schemes for DDoS mitigation, Botnets, Email spam filtering, and studying the emerging threats in new network architectures. He is a reviewer of many international conferences and journals.

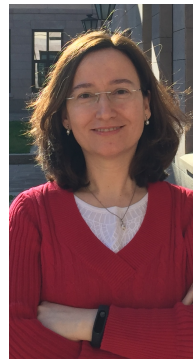

**Öznur Özkasap** is a Professor with the Department of Computer Engineering, Koç University. She received the Ph.D. degree in Computer Engineering from Ege University in 2000. She was a Graduate Research Assistant with the Department of Computer Science, Cornell University, where she completed her Ph.D. dissertation. Her research interests include distributed systems, peer-to-peer systems, energy efficiency, mobile and vehicular ad hoc networks, and security in distributed systems. Professor Özkasap is an Area Editor of the Future Generation Computer Systems journal. She is a recipient of the Turk Telekom Collaborative Research Awards, the Career Award of TUBITAK (The Scientific and Technological Research Council of Turkey), and TUBITAK/NATO A2 Ph.D. Research Scholarship Abroad, and she was awarded Teaching Innovation Grants by Koç University. She received The Informatics Association of Turkey, Prof. Aydın Köksal Computer Engineering Science Award in 2019.

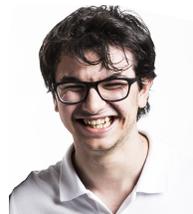

**Ahmet Uysal** is a senior research student at Department of Computer Engineering, Koç University. His research interests include computer networks and security, and computer science education. He is also interested in web technologies and web-based interactive learning tools.

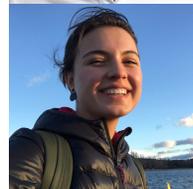

**Ceren Kocaoğullar** is a senior research student at Departments of Computer Engineering and Media & Visual Arts, Koç University. Her research interests include computer and network security, computer science education, and its technologies.

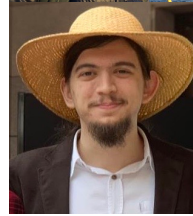

**Kaan Yıldırım** is a senior research student at Department of Computer Engineering, Koç University. His research interests include computer networks, network security, distributed systems, information security and cryptography. He is also interested in teaching computer science.